\documentclass[twocolumn,showpacs,groupedaddress]{revtex4}
\usepackage{graphicx}
\newcommand{\Tr}{\text{Tr}}
\begin{document}

\draft
\title{Geometric phases induced in auxiliary  qubits  by many-body
systems near its critical points}
\author{X. X. Yi  and W. Wang}
\affiliation{Department of Physics, Dalian University of
Technology, Dalian 116024, China}

\date{\today}

\begin{abstract}
The geometric phase induced in an auxiliary qubit by a many-body
system is calculated and discussed. Two kinds of coupling between
the auxiliary qubit and the many-body system are considered, which
lead to dephasing and dissipation in the qubit, respectively. As
an example, we consider the XY spin-chain dephasingly couple to a
qubit, the geometric phase induced in the qubit is presented and
discussed. The results show that the geometric phase might be used
to signal the critical points of the many-body system, and it
tends to zero with the parameters of the many-body system going
away from the critical points.

\end{abstract}

 \maketitle
\section{introduction}
Motivated by the fact that all realistic quantum systems are
coupled to their surrounding environments, many researchers have
paid their attention on the study of geometric phase in open
systems since
1980s\cite{garrison88,fonsera02,aguiera03,ellinas89,gamliel89,
kamleitner04,nazir02,carollo03,
marzlin04,whitney03,gaitan98,lidar05,yi05,yi06, lombardo06}. From
the perspective of application, the use of geometric phase in the
implementation of fault-tolerant quantum gates
\cite{zanardi99,jones99,ekert00,falci00} also requires the study
of geometric phase in open systems, because the system which needs
to be manipulated may decoher from a quantum superposition into
statistical mixtures due to coupling to its environment. Despite
of extensive studies that have been performed in this direction,
investigation into the effect of correlation among the
environmental systems on the induced geometric phase in  open
systems is lacking.

Geometric phases have been shown to associate with a variety of
condensed matter phenomena \cite{thouless82, resta94,
nakamura02,ryu06}. Nevertheless, their connection to quantum phase
transitions has only been given recently in \cite{carollo05,
zhu06, hamma06}, where it has been shown that the Berry's phase
can be used to signal the critical points  of the spin
chain\cite{carollo05}, the critical exponents were evaluated from
the scaling behavior of geometric phases\cite{zhu06}, and the
geometric phase can be considered  as a  topological test to
reveal quantum phase transitions\cite{hamma06}. In these works,
the geometric phase analyzed  is acquired by the ground state (or
the low lying excited states) of the many-body system. This gives
rise to the following question, whether the geometric phase
induced in an auxiliary system by the many-body system has
connection to the critical points of the many-body system. In this
paper, we shall try to answer this question by examining the
geometric phase induced in a qubit by the many-body system. This
makes our study different from that in \cite{carollo05, zhu06,
hamma06}. Two kinds of coupling between the auxiliary qubit and
the many-body system are considered. The first kind of coupling
commutes with the free Hamiltonian of the auxiliary qubit, leading
to dephasing in the qubit. Whereas the second one commutes with
the free Hamiltonian of the many-body system, which causes energy
loss (dissipation) of the qubit.

This paper is organized as follows. In Sec.{\rm II} we present a
general formulism to calculate and analyze the reduced density
matrix of the auxiliary qubit, and establish a connection of the
reduced density matrix to the critical points. As an example, we
calculate the geometric phase induced in  the auxiliary qubit
taking the XY spin chain as the many-body system  in Sec.{\rm
III}. And finally we conclude our results in Sec.{\rm IV}.

\section{general}
In this section, we will give a general   formalism to analyze the
reduced density matrix of a qubit coupled to a quantum many-body
system, and establish a connection between the reduced density
matrix of the qubit and the criticality in the quantum system. We
restrict ourself to consider the following qubit to many-body
system couplings. First we  consider the case where the coupling
conserves the energy of the qubit. Then we analyze the case where
the energy of the qubit does not conserve, but the energy of the
quantum many-body system conserves. The first case corresponds to
dephasing in the qubit, while the second kind of coupling results
in dissipation in the qubit.

Consider a qubit coupled to a quantum many-body system. The
Hamiltonian that governs the evolution of the whole system may
have the form
\begin{equation}
H=H_{\frac 1 2 }+H_m(\lambda)+H_i,\label{totalH}
\end{equation}
where $H_{\frac 1 2 }= \mu |\uparrow \rangle \langle \uparrow |,$
describes the free Hamiltonian of the qubit, $H_m(\lambda)$ stands
for the free Hamiltonian of the quantum many-body system, and
$H_i=gH_M\otimes |\uparrow\rangle\langle \uparrow | $ represents
the coupling between them. $H_M$ is an arbitrary operator of the
many-body system. It is clear that $[H_i,H_{\frac 1 2 }]=0$,
therefore the energy of the qubit conserves. The quantum system
described by $H_m(\lambda)$ undergoes a quantum phase transition
for parameter $\lambda=\lambda_c$. It is easy to show that the
time evolution operator for the whole system may be written as
\begin{equation}
U(t)=U_{\uparrow}(t)|\uparrow\rangle\langle
\uparrow|+U_{\downarrow}(t)|\downarrow\rangle\langle \downarrow|,
\label{ev1}
\end{equation}
with $U_{\uparrow}(t)$ and $U_{\downarrow}(t)$ satisfying
\begin{equation}
i\hbar\frac{\partial}{\partial
t}U_{\uparrow,\downarrow}(t)=H_{\uparrow,\downarrow}U_{\uparrow,\downarrow}(t),
\end{equation}
where
\begin{equation}
H_{\uparrow/\downarrow}=H_m+(\frac{\mu}{2}\pm\frac{\mu}{2})+(\frac
g 2 \pm \frac{g}{2})H_M.
\end{equation}
$| \uparrow\rangle$ and $| \downarrow\rangle$ are eigenstates of
$\sigma^z$ for the qubit. Having these expressions, we now show
that the off-diagonal elements of the reduced density matrix of
the qubit change drastically in the vicinity of critical points.
To this end, we assume that the qubit and the quantum many-body
system are initially independent, such that we take the following
product state as the initial state of the whole system
\begin{equation}
|\psi(0)\rangle=|\psi_{\frac 1
2}(0)\rangle\otimes|G_m\rangle,\label{inis}
\end{equation}
where $|G_m\rangle\rangle$ represent the ground state of $H_m$ and
$|\psi_{\frac 1 2 }
(0)\rangle=\cos\theta|\uparrow\rangle+\sin\theta|\downarrow\rangle$.
By a standard calculation, one obtains the reduced density matrix
of the qubit
\begin{equation}
\rho_{\frac 1 2}=\left( \matrix{ \cos^2\theta &
\cos\theta\sin\theta F(t) \cr
 \cos\theta\sin\theta F^*(t) & \sin^2\theta \cr } \right).\label{ha2}
\end{equation}
Here, $F(t)$ is defined by
\begin{eqnarray}
F(t)&=&\langle G_m|e^{-\frac{i}{\hbar}H_{\uparrow}t}|G_m\rangle ,
\end{eqnarray}
 this   is  the survival probability of the ground
state of the quantum system under the action of the Hamiltonian
$H_{\uparrow}$. The leading term $F(t)\simeq \langle
G_m|G_m^{\uparrow}\rangle$ of this equation  represents the
overlap function between two ground states $|G_m\rangle$ and
$|G_m^{\uparrow}\rangle$, where $|G_m^{\uparrow}\rangle$ denotes
the ground state of $H_{\uparrow}$. This overlapping function was
shown \cite{zanardi05} to take extremal values in the vicinity of
critical points. In fact, $F(t)$ represents the Loschmidt echo
which goes exponentially to zero as the many-body system
approaches the regions of criticality\cite{quan06}. This give us
the behavior of  $F(t)$ in the vicinity of critical points
$\lambda_c$ as $F(t)\sim|\lambda-\lambda_c|^{\nu}$, leading to a
dramatic change in the induced geometric phase near this point
$\lambda_c$. We would like to notice that the above discussions
hold for a general coupling between the spin and the quantum
system, provided the coupling is weak.

Next we turn to study the case when the energy of the qubit does
not conserve, but the energy of the many-body system conserves
instead. This implies that $[H_{\frac 1 2 },H_i]\neq 0,$ but
$[H_i,H_m]=0.$ Without loss of generality, we consider the
following Hamiltonian
\begin{eqnarray}
H&=&H_{\frac 1 2}+H_m(\lambda)+H_i,\nonumber\\
H_{\frac 1 2}&=&\Delta\sigma_x, \ \
H_i=(g_z\sigma_z+g_y\sigma_y)\otimes H_m(\lambda).
\end{eqnarray}
$\sigma_i (i=x,y,z)$ are  Pauli operators of the qubit.  The
interaction Hamiltonian $H_i$ can be rewritten as
\begin{equation}
H_i=ge^{i\gamma\sigma_x}\sigma_ze^{-i\gamma\sigma_x},
\end{equation}
with $\cos (2\gamma)=g_z/g $, and $g=\sqrt{g_z^2+g_z^2}.$ The fact
that the interaction Hamiltonian $H_i$ commutes with $H_m$ enables
us to write the time evolution operator of the whole system  as
\begin{equation}
U(t)=\sum_{n=1}^{N}U_n(t)|E_n(\lambda)\rangle\langle
E_n(\lambda)|,\label{ev2}
\end{equation}
where $\{|E_n(\lambda)\rangle\}$ are eigenstates of $H_m$ with
corresponding eigenenergies $\{ E_n=E_n(\lambda)\}$, and we assume
that they are  nondegenerate. It is easy to show that
\begin{eqnarray}
U_n(t)&=&e^{-iE_nt}\{
\cos\frac{\theta_n}{2}\cos\Omega_nt-i\sigma_x\cos\frac{\theta_n}{2}
\sin\Omega_nt\nonumber\\
&+&i\sin\frac{\theta_n}{2}\cos\Omega_nt[(\cos
2\gamma+\sin2\gamma)\sigma_z\nonumber\\
&+&(\sin2\gamma-\cos2\gamma)\sigma_y] \} .\label{ha3}
\end{eqnarray}
Here, $\cos\theta_n=\frac{\Delta}{\sqrt{g^2E_n^2+\Delta^2}},$ and
$\Omega_n=\sqrt{g^2E_n^2+\Delta^2}.$ Quantum phase transition
theory tells us that the ground state energy $E_0(\lambda)$
behaves drastically in the vicinity of the critical point
$\lambda_c$, this would reflect in $U_0(t)$ that is a function of
$E_0(\lambda)$. Having established this linkage, we claim that the
final state of the qubit
\begin{equation}
\rho_{\frac 1 2}(t)=\sum_n|c_n|^2U_n(t)\rho_{\frac 1
2}(0)U_n^{\dagger}(t),
\end{equation}
and the geometric phase acquired by this state  can signal quantum
critical points in the many-body system, provided the initial
state of the many-body system is a superposition of
$\{|E_n\rangle\}$, i.e., $|\psi_m(0)\rangle=\sum_{n=0}^N
c_n|E_n\rangle$ ($c_0\neq 0$), and at least two $c_n$ including
$c_0$ are not zero. We would like to emphasize  that the geometric
phase in this  case can not signal  the critical points, if the
many-body system is initially in the ground state with probability
1. This is because the qubit-system interaction could not excite
the many-body system in this situation. In other words, if the
couplings between the qubit and the many-body system commute with
the free Hamiltonian of the many-body system, the many-body system
would remain in its initial state if the initial state is one of
the eigenvalues of the free Hamiltonian $H_m$. Thus, the many-body
system would make no effect on the qubit during the dynamics, and
consequently the geometric phase  acquired by the qubit could not
signal the critical points of the many-body system.

\section{Geometric phases induced by the XY spin chain}
In this section, we will present an example to illustrate the
claim. Taken  a spin-chain described by the one-dimensional XY
model as the quantum many-body system, the Hamiltonian for the
whole system (spin+chain) may be given by
\begin{equation}
H=H_{\frac 1 2}+H_m+H_i, \label{h1}
\end{equation}
where
\begin{eqnarray}
H_{\frac 1 2}&=&\mu s^z,\nonumber\\
H_m&=&-2\sum_{l=1}^N((1+\gamma)s_l^x s_{l+1}^x+ (1-\gamma)s_l^y
s_{l+1}^y+\lambda s_l^z),\nonumber\\
H_i&=&4g\sum_{l=1}^N s^z s_l^z.\label{h2}
\end{eqnarray}
Here $s$ denotes  spin operator of the qubit which couples to the
chain spins $s_l$ $(l=1,...,N)$ located at the lattice site $l$.
The spins in the chain are coupled to the qubit through a constant
$g$. The time evolution operator $U(t)$ for the whole system may
be written as
$U(t)=\sum_{j=\uparrow,\downarrow}U_j(t)|j\rangle\langle j|.$
 It is easy to show that $U_{j}(t)
(j=\uparrow,\downarrow)$ satisfy $i\hbar\frac{\partial}{\partial
t}U_{j}(t)=H_{j}U_{j}(t)$ with $
H_{j}=-\sum_{l=1}^N(\frac{1+\gamma}{2}\sigma_l^x\sigma_{l+1}^x+\frac{1-\gamma}{2}\sigma_l^y\sigma_{l+1}^y
+\lambda_{j}\sigma_l^z), $ where $\lambda_{j}=\lambda\pm g$. If
the auxiliary qubits are initially in state $|j\rangle$, the
dynamics and statistical properties of the spin chain would be
govern by $H_{j}$, it takes the same form as $H_m$ but with
modified field strengths $\lambda_{j}$. The Hamiltonian $H_{j}$
can be diagonalized by a standard procedure\cite{lieb61} to be $
H_{j}=\sum_k\omega_{j,k}(\eta_{j,k}^{\dagger}\eta_{j,k}-\frac 1 2
), $ where $\eta_{j,k} (\eta_{j,k}^{\dagger})$ are the
annihilation (creation) operators of the fermionic modes with
frequency
$\omega_{j,k}=\sqrt{\varepsilon_{j,k}^2+\gamma^2\sin^2\frac{2\pi
k}{N}}, $  $\varepsilon_{j,k}=(\cos\frac{2\pi k}{N}-\lambda_{j}),
$ $ k=-N/2, -N/2+1,...,N/2-1.$ The fermionic operator $\eta_{j,k}$
was defined by the Bogoliubov transformation as, $
\eta_{j,k}=d_k\cos\frac{\theta_{j,k}}{2}-id_{-k}^{\dagger}
\sin\frac{\theta_{j,k}}{2}, $ where
$d_k=\frac{1}{\sqrt{N}}\sum_la_l \mbox{exp}(-i2\pi lk/N),$ and the
mixing angle $\theta_{j,k}$ was defined by
$\cos\theta_{j,k}=\varepsilon_{j,k}/\omega_{j,k}.$ Fermionic
operators $a_l$ were connected with the spin operators by the
Jordan-Wigner transformation via
$a_l=(\prod_{m<l}\sigma_m^z)(\sigma_l^x+i\sigma_l^y)/2$. The
  operators $\eta_{j,k}$  parameterized by
$j$ clearly do not commute with each other, this will leads to
nonzero geometric phase in the auxiliary qubits as shown later on.
Before going on to calculate the reduced density matrix, we
present a discussion on the diagonalization of  $H_{j}$. For a
chain with periodic boundary condition, i.e., $\sigma_1=\sigma_N,$
boundary terms $H_{boun}\sim [(a^{\dagger}_Na_1+\gamma
a_Na_1)+h.c.][exp(i\pi M)+1]$ have to be taken into
acount\cite{lieb61, katsura62}.   In this paper, we would work
with $H_{boun}$ ignored, because we are interested in finding a
link between the criticality of the chain and the entanglement in
the auxiliary qubits. In fact, the boundary effect can be ignored
provided $N\rightarrow \infty$, this is exactly the situation we
consider in this paper.

Having given an initial product(separable) state of the total
system, $|\psi(0)\rangle=$ $ |\psi_{\frac 1 2}(0) \rangle$ $
\otimes|\phi_m(0)\rangle,$ we can obtain the reduced density
matrix for the auxiliary qubits as $\rho_{\frac 1 2}(t)=$
$\Tr_m[U(t)|\psi(0)\rangle\langle\psi(0)|U^{\dagger}(t)],$ it may
be formally written in the form
\begin{equation}
\rho_{\frac 1 2}(t)=\sum_{i,j}\rho_{ij}(t)|i\rangle\langle
j|.\label{stateab0}
\end{equation}
A straightforward but somewhat tedious calculation shows that
$\rho_{ij}(t)=\rho_{ij}(t=0)F_{ij}(t),$ with
\begin{widetext}
\begin{eqnarray}
F_{ij}(t)&=&\prod_ke^{\frac i 2 (\omega_{i,k}-\omega_{j,k})t}
\{1-(1-e^{i\omega_{i,k}t})\sin^2\frac{\theta_k-\theta_{i,k}}{2}-
(1-e^{-i\omega_{j,k}t})\sin^2\frac{\theta_k-\theta_{j,k}}{2}\nonumber\\
&+&(1-e^{i\omega_{i,k}t})
(1-e^{-i\omega_{j,k}t})[\sin\frac{\theta_k-\theta_{i,k}}{2}\sin\frac{\theta_k-\theta_{j,k}}{2}
\cos\frac{\theta_{i,k}-\theta_{j,k}}{2}]\}, \label{stateab}
\end{eqnarray}
\end{widetext}
where $\cos\theta_k=\frac{\cos(2\pi k/N)-\lambda}{\sqrt{(\cos
(2\pi k/N)-\lambda)^2+\gamma^2\sin^2(2\pi k/N)}}.$ To derive this
result, the spin chain was assumed to be initially in the ground
state of $H_m$.

With these expressions, we now turn to  study the geometric phase
of  the open system. For an open system, the state in general  is
not pure and the evolution of the system is not unitary. For
non-unitary evolution as shown in Eq.(\ref{stateab}), the
geometric phase can be calculated as follows. First, solve the
eigenvalue problem for the reduced density matrix $\rho_{\frac 1
2}(t)$ and obtain its eigenvalues $\varepsilon_k(t)$ as well as
the corresponding eigenvectors $|\psi_k(t)\rangle;$ Secondly,
substitute $\varepsilon_k(t)$ and $|\psi_k(t)\rangle$ into
\begin{widetext}
\begin{equation}
\Phi_g= \mbox{arg}(\sum_k\sqrt{\varepsilon_k(0)\varepsilon_k(T)}
\langle\psi_k(0)|\psi_k(T)\rangle
e^{-\int_0^T\langle\psi_k(t)|\partial/\partial t|\psi_k(t)\rangle
dt}).\label{gp5}
\end{equation}
\end{widetext}
Here, $\Phi_g $ is the geometric phase for the system undergoing
non-unitary evolution \cite{tong04}, $T$ is the total evolution
time. The geometric phase Eq. (\ref{gp5}) is gauge invariant and
can be reduced to the well-known results in the unitary evolution.
It is experimentally testable. The geometric phase factor defined
by Eq.(\ref{gp5}) may be understood as a weighted sum over the
phase factors pertaining to the eigenstates of the reduced density
matrix, thus the detail of analytical expression for the geometric
phase would depend on the digitalization of the reduced density
matrix Eq.(\ref{stateab}). The eigenvalues of the reduced density
matrix are readily calculated
\begin{equation}
\varepsilon_{\pm}(t)=\frac 1 2 \pm \frac 1 2
\sqrt{1-4\rho_{11}\rho_{22}+4|\rho_{12}|^2}.
\end{equation}
In order to calculate the geometric phase, we only need the
eigenvalue $\varepsilon_+(t)$ and its corresponding eigenvector
\begin{equation}
|\varepsilon_+(t)\rangle=\cos\frac{\alpha}{2}e^{-i\phi}|\uparrow\rangle+
\sin\frac{\alpha}{2}|\downarrow\rangle,
\end{equation}
since $\varepsilon_-(0)=0.$ Here
$\cos\alpha=\frac{(\rho_{11}-\rho_{22})}{2}/\sqrt{|\rho_{12}|^2+\frac
1 4 |\rho_{11}-\rho_{22}|^2},$ and $\tan\phi={\mbox Im}
(\rho_{12})/{\mbox Re}(\rho_{12}),$ i.e., the ratio of the
imaginary part   to the real part of $\rho_{12}.$ Having these
results, we obtain the geometric phase,
\begin{equation}
\Phi_g=\mbox{arg}( \sqrt{\varepsilon_+(0)\varepsilon_+(T)}
\langle\varepsilon_+(0)|\varepsilon_+(T)\rangle
e^{i\int_0^T\cos^2\frac{\alpha}{2} \dot{\phi} dt}).\label{gp6}
\end{equation}
We would like to notice that the geometric phase vanishes with
$\gamma=0$ because $F_{ij}(t)=1$ at this point. $F_{ij}(t)=1$
results in $\phi=0$ in Eq.(\ref{gp6}), this together with that
$\langle\varepsilon_+(0)|\varepsilon_+(T)\rangle$ is real yield
zero geometric phase at this critical point.
\begin{figure}
\includegraphics*[width=0.9\columnwidth,
height=0.75\columnwidth]{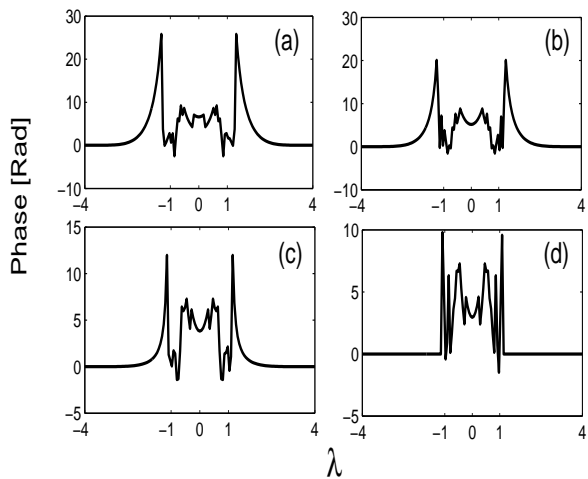} \caption{The geometric
phase as a function of   $\lambda$. The figure was plotted for
$N=1245$ sites, $g=0.1$ and
(a)$\gamma=0.2$,(b)$\gamma=0.5$,(c)$\gamma=0.8$,(d)$\gamma=1$.}
\label{fig1}
\end{figure}

Now we turn to study the criticality in the transverse Ising
model($\gamma=1$ in the XY model). The ground state structure of
this model change drastically as the parameter $\gamma$ is varied.
We first summarize the ground states of this model in the limits
of $|\lambda|\rightarrow \infty$, $|\lambda|=1$ and $\lambda=0$.
The ground state of the spin-chain approaches a product of spins
pointing the positive/negative  $z$ direction in the
$|\lambda|\rightarrow \infty$ limit, whereas the ground state in
the limit $\lambda=0$ is doubly degenerate under the global spin
flip by $\prod_{l=1}^N\sigma_l^z$. At $|\lambda|=1$, a fundamental
transition in the ground state occurs, the symmetry under the
global spin flip breaks at this point and the chain develops a
nonzero magnetization $\langle\sigma_x\rangle\neq 0$ which
increases with $\lambda$ growing. The above mentioned properties
of the ground state are reflected in the geometric phase as shown
in figure \ref{fig1}-(d). In the limit $|\lambda|\rightarrow
\infty$, $\theta_{j,k}=\theta_k=\pi/(-\pi)$,
 this results in $\Phi_g=0$. In fact  when
$|\lambda|\geq 4$, $\Phi_g$ approaches zero very well. Around
$|\lambda| \rightarrow 1$, the geometric phase changes
drastically, this can be interpreted as the sensitivity of the
spin-chain ground state to perturbations from the qubit-chain
coupling at these points. The difference between $\gamma=0$ and
$\gamma=1$ is that $F_{ij}(t)=1$ for $\gamma=0$, but it does not
hold for $\gamma=1$. This is the reason why the geometric phase
takes different values at these critical points, as figure
\ref{fig1} shows. The ground state of the XY model is really
complicated with many different regime of
behavior\cite{barouch70}, these are reflected in sharp changes in
the geometric phase across the line $|\lambda|=1$ regardless of
$\gamma$ (as shown in figure 1-(a),(b) and (c)), indicating the
change in the ground state of the spin-chain from paramagnetic
phase to the others. Instead of discussing the scaling behavior of
the induced geometric phase, we consider the scaling property of
$F_{ij}(t)$ given in Eq.(\ref{stateab}) in the vicinity of the
critical points. Noticing that $F_{ij}(t)$ is a function of
$(\theta_k-\theta_{j,k})$ $(j=\uparrow,\downarrow)$, the scaling
behavior of $F_{ij}(t)$ may be characterized by
$S_N^{\lambda}(\lambda,\gamma)=\sum_{k=1}^M(\frac{\partial
\theta_k}{\partial \lambda})^2$, and
$S_N^{\gamma}(\lambda,\gamma)=\sum_{k=1}^M(\frac{\partial
\theta_k}{\partial \gamma})^2$\cite{zanardi05}. It was shown that
$S_N^{\lambda}(|\lambda|=1,\gamma)\sim N^2/\gamma^2$ and
$S_N^{\lambda}(|\lambda|\leq 1,\gamma=0)\sim N^2$  in the vicinity
of criticality  for the one-dimensional XY Model.

\section{conclusion}
In conclusion, by discussing the geometric phase in the auxiliary
qubit coupled to   the many-body systems, the relation between the
geometric phase induced in the qubit and the critical points of
the many-body system  was established.  The induced geometric
phase in the qubit change drastically at the critical points, this
is due to the sensitivity of the many-body systum to parameter
changes near its critical points. The relation not only provides
an efficient theoretical tool to study quantum phase transitions,
but also proposes a method to measure the critical points in
experiments\cite{note1}. The limitation of this discussion is that
the coupling between the qubit  and the quantum system is assumed
weak, and as we have shown, the geometric phase could not reflect
the critical points of the quantum system that is initially in its
ground state
 with qubit-system coupling satisfying $[H_m, H_i]=0$.

\ \ \\
This work was supported
by NCET of M.O.E, and NSF of China under Grant  No. 60578014.\\


\begin{references}
\bibitem{note1} The experiment to observe the prediction in this paper
can be envisioned as follows. First let the qubit (a two-level
system) transmit through a 50/50 beam splitter, the qubit is split
and follows two different paths then. On one of the paths, the
many-body system is placed and interact with the qubit. An
interference patten can be observed when the two paths meet, this
is similar to the Mach-Zehnder interferometer.
\end{references}
\end{document}